# Dielectric response of the charge-ordered two-dimensional nickelate $La_{1.5}Sr_{0.5}NiO_4$

J. Rivas[1], B. Rivas-Murias[1], A. Fondado[1], J. Mira[1] & M. A. Señarís-Rodríguez[2]

[1] Departamento de Física Aplicada, Universidade de Santiago de Compostela, E-15782 Santiago de Compostela, Spain

[2] Departamento de Química Fundamental, Universidade da Coruña, E-15071 A Coruña, Spain

**Abstract**

We are reporting the dielectric response of $La_{1.5}Sr_{0.5}NiO_4$, a system that presents a charge-ordered state above room temperature and a rearrangement of its charge-order pattern in the temperature region 160-200 K. A careful analysis of the role of the electric contacts used, sample thickness and grain size on the experimental data allows us to determine that this material exhibits high values of intrinsic dielectric constant. The variation of the dielectric constant with temperature shows a maximum in the region of the rearrangement of the charge-order pattern, which constitutes an evidence of the link between both phenomena.

Ferroelectrics are the standard solution to obtain devices that make use of notable dielectric properties. In them, the ferroelectric state arises because the centers of positive charges in the crystalline lattice do not coincide with those of negative charges. The ultimate origin of the ferroelectric state is then, structural. Among the alternative strategies to find materials with high dielectric constant are those that involve condensation of electronic charges. The interest on these alternatives has grown since the report by Homes *et al.*[1] of a high-dielectric constant of about $8x10^4$ in $CaCu_3Ti_4O_{12}$ at temperatures as high as 250 K and frequencies up to 1 MHz. The origin of this finding has been discussed by Lunkenheimer *et al.*[2,3], who attribute it to extrinsic effects. Basically, they argue that the temperature independent dielectric constant in a broad temperature range found by Homes *et al.* cannot stem from permanent dipoles or off-center ions. He *et al.*[4] and Cohen *et al.*[5] also attribute to extrinsic effects the dielectric response of $CaCu_3Ti_4O_{12}$, after a first-principles study of its structural and lattice dielectric response. In the same line, Sinclair *et al.*[6], stress the apparent character of this colossal dielectric constant and explain it by Maxwell-Wagner-type contributions of depletion layers at the interface between sample and contacts and at grain boundaries.

Despite this controversy, it seems admissible to think about a correlation between the dielectric properties and the electronic state. In this context, we have focused on systems with charge condensation, starting with charge-ordered manganese perovskites. Our report on the finding of a high capacitive behaviour in $Pr_{2/3}Ca_{1/3}MnO_3$ that appears just below its charge ordering temperature, $T_{CO}$ = 250 K (refs. 7, 8), is an evident demonstration of the link between the electronic state and the increase of the dielectric response.

In this letter we are focusing on a system with charge ordering temperature above room temperature, the mixed oxide $La_{1.5}Sr_{0.5}NiO_4$, with $T_{CO}$= 480 K (ref. 9), with the aim of getting a high dielectric constant at ambient conditions. In order to test the role of intrinsic and extrinsic factors, and on the basis of refs. 2-6, the sample contacts were changed, the sample thickness was modified, and two kinds of samples were synthesized by two routes giving different grain sizes. On one hand, a $La_{1.5}Sr_{0.5}NiO_4$ ceramic sample was prepared by a conventional solid-state reaction, starting from stoichiometric amounts of dry $La_2O_3$, $SrCO_3$ and NiO, that were thoroughly mixed and grinded together, pressed into pellets and fired at 1373, 1473 and 1573 K with intermediate grindings. The sample was then cooled to room temperature at the rate of 1 K/min. On the other hand, the same compound was synthesized by the Pechini method using $La_2O_3$, $SrCO_3$ and $Ni(NO_3)_2 \cdot 6H_2O$ as starting materials. The procedure was as follows: $La_2O_3$ was first converted into the corresponding nitrate by dissolution in 30% nitric acid. This product was then added to a 1M citric acid aqueous solution, in which stoichiometric amounts of $SrCO_3$ and $Ni(NO_3)_2 \cdot 6H_2O$ were also dissolved. After diluting the so-obtained solution, we carefully added ethyleneglycol in a proportion of 10% v/v. The resulting solution was heated at 473 K until we obtained a brown resin, whose organic matter was subsequently decomposed at 673 K. The obtained ashes were given accumulative heating treatments at 873, 973, 1073 and 1173 K followed by intermediate grindings. The pelletized sample was finally annealed at 1273 K/48 hours.

Both samples were characterized by X-ray powder diffraction, that showed that they are single phase, with a structure related to the perovskite: $La_{1.5}Sr_{0.5}NiO_4$ displays a quasi two-dimensional structure ($K_2NiF_4$) in which perovskite blocks, that are one-octahedra thick, are separated from one another by the presence of rock-salt type (La-Sr/O) layers along the c-axis. Also, by scanning electronic microscopy it was observed that the obtained polycrystalline materials have an averaged particle diameter of 0.7 μm for the Pechini sample and several micrometers for the ceramic sample. The complex dielectric permittivity was measured with a parallel-plate capacitor coupled to a precision LCR meter Agilent 4284 A, capable to measure in frequencies ranging from 20 to $10^6$ Hz. The capacitor was mounted in an aluminium box refrigerated with liquid nitrogen, and incorporating a mechanism to control the temperature. The samples were prepared to fit in the capacitor, and gold and silver were sputtered on their surfaces to ensure good electrical contact with the electrodes of the capacitor. The system was tested using a commercial $SrTiO_3$ sample, and gave values similar to those reported in the literature[10].

The complex relative dielectric permittivity of $La_{1.5}Sr_{0.5}NiO_4$,

$$\varepsilon_r(\omega) = \varepsilon_r'(\omega) - i\varepsilon_r''(\omega) \qquad (1)$$

($\varepsilon_r = \varepsilon / \varepsilon_0$ ; where $\varepsilon_0 = 8.85\mathrm{x}10^{-12}$ F/m is the permittivity of free space and $\omega$ is the angular frequency) was measured as a function of frequency and temperature. In Fig. 1a we show the real part of the relative permittivity (dielectric constant), $\varepsilon'_r$, of the ceramic sample with sputtered gold contacts, in the frequency range from 20 Hz to 1 MHz at several temperatures. It is higher than $10^6$ at very low frequencies, keeps well above $10^5$ at room temperature up to 100 kHz and decreases to $3\mathrm{x}10^4$ at 1 MHz.

The same measurement was done in the sample of smaller grain size, and the values changed (Fig 1a), indicating a dependence on the grain size. Additionally, it was observed that sample contacts play also a certain role as it can be seen in Fig 1b. Finally, the thickness of the sample was changed and it was observed that it altered the dielectric response (Fig 1c). It is clear therefore the existence of all these extrinsic factors in the determination of the dielectric constant of $La_{1.5}Sr_{0.5}NiO_4$.

The frequency dependence of the imaginary part, $\varepsilon''_r$ was also measured in the Pechini sample (Fig. 2a). In order to study frequency dependent or purely ac relaxation effects, it is better to substract the dc contribution account from the $\varepsilon''_r$ observed value, taking into account that[11]

$$\varepsilon''_{r,die}(\omega) = \varepsilon''_r(\omega) - \frac{\sigma_{dc}}{\varepsilon_0\omega} \qquad (2)$$

($\varepsilon''_{r,\,die}$ = loss-factor due to a true dielectric response, $\sigma_{dc}$= dc electric conductivity). Then, we can substract to the data of Fig. 2a the contribution of migrating charge carriers, $\sigma_{dc}/(\varepsilon_0\,\omega)$, where the dc conductivity, $\sigma_{dc}$, is obtained from the extrapolation at low frequencies of the conductivity, $\sigma(\omega)$, shown in Fig. 2b. We obtain in this way the result presented in Fig. 2c, that enables us to observe the evolution with frequency and temperature of the dielectric relaxation, and to obtain the characteristic frequency of the relaxation with increasing temperature. There is a noticeable increase of the characteristic relaxation times, $\tau=1/\omega$, with decreasing temperatures. A logarithmic fit of the characteristic times versus inverse temperature shows two different regimes with Arrhenius behaviour, $\tau = \tau_0 \exp(U/(k_B T))$, where U is the activation energy and $k_B$ the Boltzmann constant, (Fig 2d): one at high temperatures (>200 K) with activation energy U~73 meV and another at lower temperatures with U~51 meV.

If we examine the temperature dependence of $\sigma_{dc}$ we observe a thermal activated behaviour, with activation energy ~72 meV, in the high temperature regime (Fig. 2e). This behaviour changes in the region 160-200 K, to another with activation energy of ~44 meV. This situation correlates well with that of characteristic times (Fig 2d), establishing a link between its dielectric relaxation and the conductivity. What is the reason for this change at 160-200 K? We think that it could be found in a recent work by Kajimoto *et al.*[9], who have studied in detail the charge ordering of $La_{1.5}Sr_{0.5}NiO_4$ by neutron diffraction: they found, at 180 K, a spontaneous rearrangement of such charge ordering, from a checkerboard pattern to a stripe-type charge order. This rearrangement does not only affect the activation energies commented before, but also the high dielectric constant itself, that falls below this temperature (Fig. 3).

From the results of Fig. 1 we observe the influence of extrinsic factors, like sample contacts, thickness and grain size. The dielectric spectra reveal a quasi-Debye relaxation which can be explained satisfactorily with the help of a two- or tri-layer

Maxwell-Wagner capacitor.[12] The dispersion of dielectric constant can be modelled taking into account the interfacial polarization due to the existence of depletion layers near the Au/Ag contacts, and the polycrystalline solid can be imagined to consist of well-conductive grains separated by layers of lower conductivity.[6] In this phenomenological model, if we assume that the true dielectric constant is almost the same in all the sample (in the contacts zone and grain boundaries the conductivity of the material changes, but the dielectric constant of the material should not be altered too much), then it can be demonstrated[12] that the measured dielectric constant at its optical value ($\omega \rightarrow \infty$), $\varepsilon'_\infty$, is the intrinsic value of the material in question. Although this optical value is in the limits of our experimental device, we have estimated a value around $\varepsilon'_\infty \approx 40$ at room temperature for the intrinsic dielectric constant of this nickelate. In a Mawxell-Wagner approach, the $\varepsilon'_r$ high values observed can be explained using this intrinsic value, multiplied by a term dependent on the aforementioned extrinsic factors.

The role of the intrinsic part is manifested in the thermal dependence of the measured dielectric constant (Fig. 3). Given that the conductivity increases monotonically with temperature (Fig. 2b), the extrinsic contribution also changes monotonically with temperature, and therefore the thermal dependence of the dielectric constant must be shaped by the intrinsic term. The maximum of the measured dielectric constant is found around 160-200 K (indicating that the intrinsic term should peak in a similar temperature range), a temperature region where, as noted before, $La_{1.5}Sr_{0.5}NiO_4$ undergoes a spontaneous rearrangement of its charge ordering pattern[9]. Then there seems to be a link between the electronic state of the material and its dielectric function.

In summary, we believe that this work is of general interest due to: (i) it reports rather high values of the dielectric constant in $La_{1.5}Sr_{0.5}NiO_4$ (compared to most other technically relevant materials, as pointed our by Lunkenheimer *et al.*[3]); (ii) it shows a correlation with the charge-order pattern of the material, indicating a link of both phenomena; (iii) from the applied point of view, it proposes new work to be done in other charge-ordered compounds, more suitable to minimize the dielectric losses, while keeping $\varepsilon'_r$ as high as possible.

We thank the financial support of the DGI of the Ministry of Science and Technology of Spain under project FEDER MAT2001-3749. We also acknowledge Dr. Francisco Rivadulla (University of Santiago de Compostela) for a critical reading of this manuscript.

Figure 1. (a) Real part of the complex relative dielectric permittivity, $\varepsilon'_r$, of $La_{1.5}Sr_{0.5}NiO_4$ versus frequency at selected temperatures, for two grain sizes. Data taken in a commercial $SrTiO_3$ sample are included as reference. The same measurement, in the ceramic sample, with two kinds of contacts (b) and two different sample thicknesses (c).

Figure 2. In a $La_{1.5}Sr_{0.5}NiO_4$ sample synthesized by the Pechini method: (a) Imaginary part of the complex relative dielectric permittivity, $\varepsilon''_r$, versus frequency at selected temperatures, measured with gold contacts. (b) Conductivity vs. frequency at diverse temperatures Extrapolation of the curves to zero frequency gives $\sigma_{dc}$. (c) Frequency dependence of the imaginary part of the complex relative dielectric permittivity after substraction of the contribution of free charge carriers. The maxima define the characteristic frequencies. (d) Logarithm of the characteristic times vs. the inverse of temperature. From linear fits we obtain two activation energies, with a boundary around 200 K. (e) Logarithm of $\sigma_{dc}$ vs. the inverse of temperature. Above 200 K the linear fit gives an activation energy of ~70 meV.

Figure 3. $\varepsilon'_r$ vs. temperature at selected frequencies in the Pechini sample. The maxima separate two different behaviours of the dielectric response of $La_{1.5}Sr_{0.5}NiO_4$.

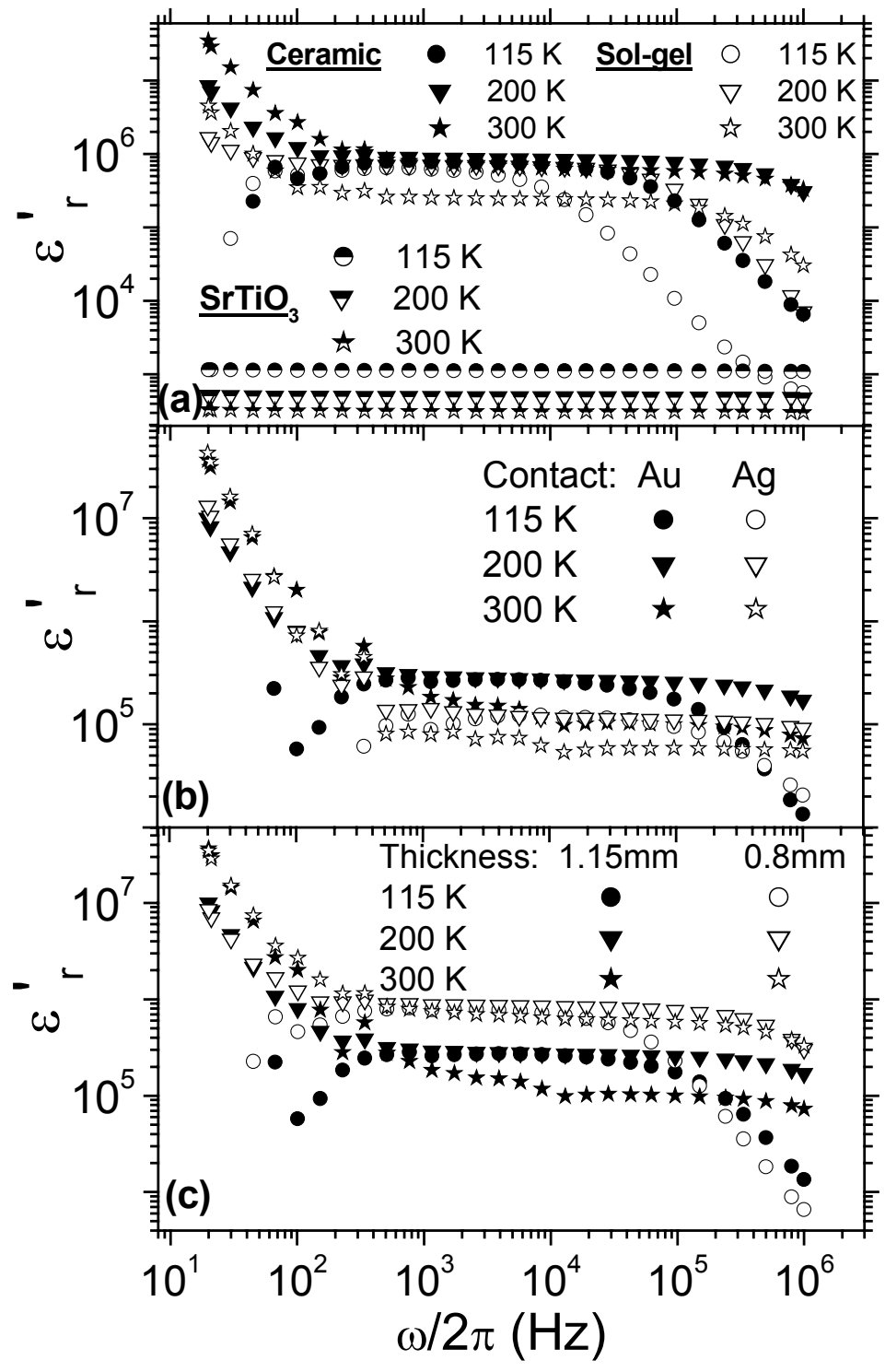

Figure 1

J. Rivas et al.

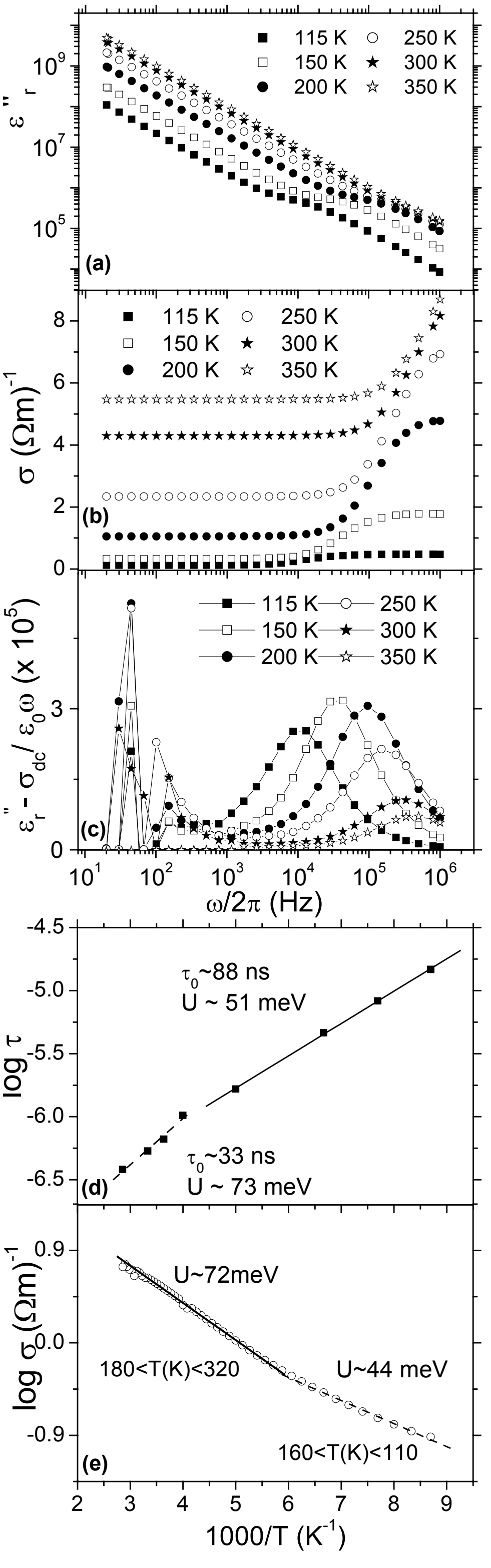

Figure 2 J. Rivas et al.

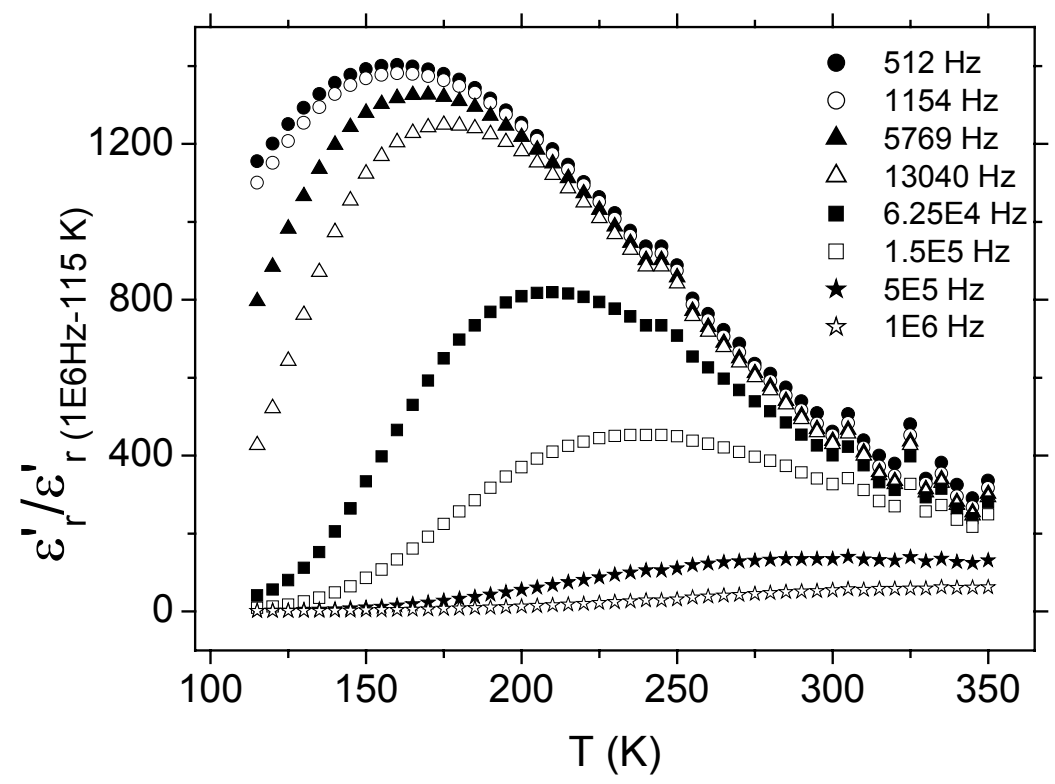

Figure 3 J. Rivas et al.